# Vulnerability Detection Using Two-Stage Deep Learning Models

Mohamed Mjd Alhafi*, Mohammad Hammade and Khloud Al Jallad

*Department of Information and Communication Engineering, Arab International University, Daraa, Syria.*

**\*Corresponding author**
Mohamed Mjd Alhafi, Department of Information and Communication Engineering, Arab International University, Daraa, Syria.

## Abstract
*Application security is an essential part of developing modern software, as lots of attacks depend on vulnerabilities in software. The number of attacks is increasing globally due to technological advancements. Companies must include security in every stage of developing, testing, and deploying their software in order to prevent data breaches. There are several methods to detect software vulnerability Non-AI-based such as Static Application Security Testing (SAST) and Dynamic Application Security Testing (DAST). However, these approaches have substantial false-positive and false-negative rates. On the other side, researchers have been interested in developing an AI-based vulnerability detection system employing deep learning models like BERT, BLSTM, etc. In this paper, we proposed a two-stage solution, two deep learning models were proposed for vulnerability detection in C/C++ source codes, the first stage is CNN which detects if the source code contains any vulnerability (binary classification model) and the second stage is CNN-LTSM that classifies this vulnerability into a class of 50 different types of vulnerabilities (multiclass classification model). Experiments were done on SySeVR dataset. Results show an accuracy of 99% for the first and 98% for the second stage.*

## Introduction
Attacks are defined as the exploitation of vulnerabilities in a system, where these vulnerabilities are caused by human errors with no malicious intent. The attacker relies heavily on these vulnerabilities when performing the hack in order to damage the system. In general, vulnerabilities are the main gateway for the occurrence of combustion. With the massive development in technology in the last decades, electronic attacks pose a real danger to systems especially banks. Therefore, it is necessary to have tools that help developers detect vulnerabilities in their source codes, in order to make their system more secure and not contain any gateway that allows attackers to enter the system. Detect vulnerabilities techniques are either AI-based or non-AI-based such as Static application security testing (SAST) and dynamic application security testing (DAST).

## Related Works
### Datasets
Many datasets were created for vulnerability detection. We will list some of them. The Software Assurance Reference Dataset (SARD)( is a growing collection of over 170 000 programs with precisely located bugs [1]. The programs are in C, C++, Java1, PHP, and C# and cover more than 150 classes of vulnerabilities. The National Vulnerability Database (NVD)) is a publicly accessible data repository that keeps standardized data on reported software vulnerabilities [2]. More than 43,000 software vulnerabilities impacting more than 17,000 software programs have been disclosed since NVD's founding in 1997. CVEFixes is a comprehensive vulnerability dataset that is automatically collected and curated from Common Vulnerabilities and Exposures (CVE) records in the public [3]. CVEFixes dataset is structured as a relational database consisting of multiple tables where each table presents artifacts at each specific abstraction level. The ManySStuBs4J is a collection of quick patches for Java problems called corpus is a tool for assessing software repair methods [4]. All bug-fixing changes are collected using the SZZ heuristic, which is then filtered to provide a data set of minor bug-fix changes These single statement fixes are categorized into one of 16 syntactic templates that invoke SStuBs whenever it is practical. Simple statement defects from open-source Java projects posted on GitHub are included in the collection. There are two variants of the dataset. One was taken from the top 1000 Java Projects, and the other was taken from the top 100 Java Maven Projects. The ESC consists of 40,932 smart contracts from Ethereum with roughly 307,396 functions in total [5].

Approximately 5,013 of the functions have at least one callable invocation. They may be impacted by the reentrancy vulnerability because of their potential value. Approximately 4,833 functions make up the block. They are vulnerable to the timestamp reliance issue because they use timestamp statements. Decentralized blockchain platform Ethereum may be used to create a variety of applications. The VSC the dataset includes all 4,170 smart contracts that were gathered from the VNT Chain network and total 13,761 functions [6]. An experimental public blockchain network called VNT Chain has been put up by organizations and academic institutions in Singapore, China, and Australia. In They collected the Semantics-based Vulnerability Candidate (SeVC) dataset, which includes every type of vulnerability that is accessible through the National Vulnerability Database (NVD) and the Software Assurance Reference Dataset (SARD) [7]. The SeVC dataset focuses on 14,000 SARD applications and 1,591 open-source C/C++ programs from the NVD. There are 420,627 SeVCs total, of which 56,395 SeVCs are vulnerable and 364,232 SeVCs are not. Four types of SyVCs are involved: Library/API Function Call, Array Usage, Pointer Usage, Arithmetic Expression.

| Name | #Samples | Programming Language | Label Type | Year |
|---|---|---|---|---|
| CVEFixes [3] | 61K | C/C++ | Binary | 2021 |
| SySeVR [7] | 420K | C/C++ | Multiple class | 2021 |
| ReVeal FFMPeg+Qemu [8] | 40K | C | Binary | 2020 |
| ManySStuBs4J [4] | 63K | Java | Multiple class | 2019 |
| Drapper VDISC [9] | 1.27M | C/C++ | Multiple class | 2018 |

**Table 1: Public Datasets Statistics**

## Previous Works

Most recent studies have focused on the use of deep learning, since the source code can be represented as a graph, there are researchers who have applied Graph Neural Network (GNN), because the structure of the graph can capture the decencies and relation among nodes a lot of researches use graph as in [8-10]. While other studies found that the source code is a text where natural language processing techniques can be used to extract features from it, such as Bag of Words [11-13]. Saikat Chakraborty et al. in used Code Property Graph to represent the original code, then they use GGNN that assigns a Gated Recurring Unit (GRU) to update the current vertex embedding by assimilating the embedding of all its neighbors [8]. Zhen Lis et al. In represented program as vector using the notation of Syntax-based Vulnerability Candidates (SyVCs) and Semantics-based Vulnerability Candidates (SeVCs) depending on Abstract Syntax Tree (AST) [9]. Anshul Tanwa et al. in used Abstract Syntax Tree (AST) to extract features from the code then embedded the features vector to fit them in encoder that consists of self-attention module, BLSTM and Conv layer [10]. Rebecca L. Russell et al. in proposed using (BOW) because it is simple representing text data and has seen great success in problems such as language modeling and document classification [12]. In order to represent programs in vectors that are suitable for the input to neural networks, in proposed transforming programs into a representation of code gadget, which is composed of a number of program statements (i.e., lines of code), which are semantically related to each other in terms of data dependency or control dependency [11].

| Paper | Dataset | Model | F1 | Year |
|---|---|---|---|---|
| SySeVR: A Framework for Using Deep Learning to Detect Software Vulnerabilities [10] | SySeVR [7] | LR | 0.62 | 2021 |
| | | CNN | 0.81 | |
| | | MLP | 0.68 | |
| | | DBN | 0.63 | |
| | | LSTM | 0.79 | |
| | | GRU | 0.81 | |
| | | BLSTM | 0.83 | |
| | | BGRU | 0.83 | |
| Combining Graph Neural Networks with Expert Knowledge for Smart Constract Vulnerablility Detection [11] | ESC & VSC [5] | RNN | 0.45 | 2021 |
| | | LSTM | 0.54 | |
| | | GRU | 0.54 | |
| | | GCN | 0.71 | |
| | | CGE | 0.87 | |
| Deep Learning based Vulnerability Detection: Are We There Yet? [12] | ReVeal [8] | RF | 0.25 | 2020 |
| | | MLP | 0.36 | |
| | | SVM | 0.39 | |
| | | GRU | 0.41 | |
| | FFMpeg + Qemu [8] | RF | 0.52 | |
| | | MLP | 0.59 | |
| | | SVM | 0.60 | |
| | | GRU | 0.64 | |

| Automated Vulnerability Detection in Source Code Using Deep Representation Learning [13] | Draper VDISC [9] | BOW+RF | 0.78 | 2018 |
| --- | --- | --- | --- | --- |
| | | RNN | 0.80 | |
| | | CNN | 0.84 | |
| | | CNN+RF | 0.82 | |
| | | RNN+RF | 0.81 | |
| A Deep Learning-Based System for Vulnerability Detection [12] | CVEFixes [3] | BLSTM | 0.95 | 2018 |

**Table 2: Previous Works Results**

## Proposed Solution

Our proposed solution considered source code from the SySeVR dataset as a raw text where tokenization was used to convert text into a numeric context to train a deep learning model [7]. But tokenization has a drawback because it does not capture the semantic dependency between different code lines. Because it assigns an ID to each unique word far from the semantic meaning of the word. So, because the programming language is a context-free grammar, the variable namespace is open. This leads to an increase in the number of tokens, even if the tokens have the same role. Therefore, regular expressions were used in order to identify the names of the variables and functions to convert them into fixed tokens. For example, if the first source code has a variable "X" and the second source code has a variable "Y". In this case, the tokenizer will assign an ID for the token "X" different than the token "Y". But "X" and "Y" are variables and they do the same role in the programming language. If they are replaced by a constant word such as "VAR" for variables and "FUNC" for functions, in all samples, the first variable will be named "VAR1" and the second will be named "VAR2" etc. (see Figure 1). In this way, the tokenizer gives the first variable in all dataset samples the same ID. Thus, we have preserved the semantic dependency in the source codes.

**Figure 1:** Algorithm to Obtain Semantic Dependency from the Input Code

```
Algorithm 1: Algorithm
Input: Code
Output: Processed Code
tokens ← getTokens(Input)
idVar ← 0
idFunction ← 0
repeat
  for each: t ∈ tokens
    Do this
    if t = Digit then
      t ← "NUMBER"
    if t = String then
      t ← "STRING"
    if t = Variable then
      t ← "VAR" + idVar
      idVar ← idVar +1
    if t = FunctionName then
      t ← "FUNC" + idFunction
      idFunction ← idFunction+1
return Tokens
```

| Sample Type | #Samples |
| --- | --- |
| Non-vulnerable | 88K |
| Vulnerable | 30K |

**Table 3: Samples Per Class in SySeVR Dataset**

There are approximately 88k samples that do not contain vulnerable. While on the vulnerable samples the class with the most samples has only 4930 samples (CWE121).we proposed a solution to solve this problem by dividing the main task into two tasks, each task is done by a model, by using the dataset in different ways. , first stage uses the whole samples and the model trained over it to detect if the source code has vulnerable or not (Binary classification), while the second stage trained on dataset that contains only vulnerable samples to detect the type of this vulnerable (Multiclass classification). Synthetic Minority Oversampling Technique (SMOTE) was used to overcome imbalance problem, SMOTE first selects a minority class instance at random and finds its k nearest minority class neighbors. Then, to create the synthetic instance, a line segment in the feature space is made by randomly picking one of the k nearest neighbors b, and connecting it to a. The synthetic instances are generated as a convex combination of the two chosen instances a and b [11].

| CWE | description |
| --- | --- |
| CWE-79 | Improper Neutralization of Input During Web Page Generation (Cross-site Scripting) |
| CWE-119 | Improper Restriction of Operations within the Bounds of a Memory Buffer |
| CWE-20 | Improper Input Validation |
| CWE-125 | Out-of-bounds Read |
| CWE-200 | Exposure of Sensitive Information to an Unauthorized Actor |
| CWE-787 | Out-of-bounds Write |
| CWE-476 | NULL Pointer Dereference |
| CWE-264 | Permissions Privileges and Access Controls |

**Figure 2: CWE IDs with Description [12]**

### Stage1 (Binary Classification Model)
The model consists of one embedding layer, two 1-D convolution layers, 3 fully connected layers. The input to the network is a preprocessed text, where each word embedded into 13-D vector by Embedding Layer, and the total number of words are truncated into 500 words. The number of layers was selected so as to maintain a high level of accuracy while still being fast enough for real-time purposes. In addition, it utilized max pooling, too ) see Figure 3 Stage1 Model Architecture).

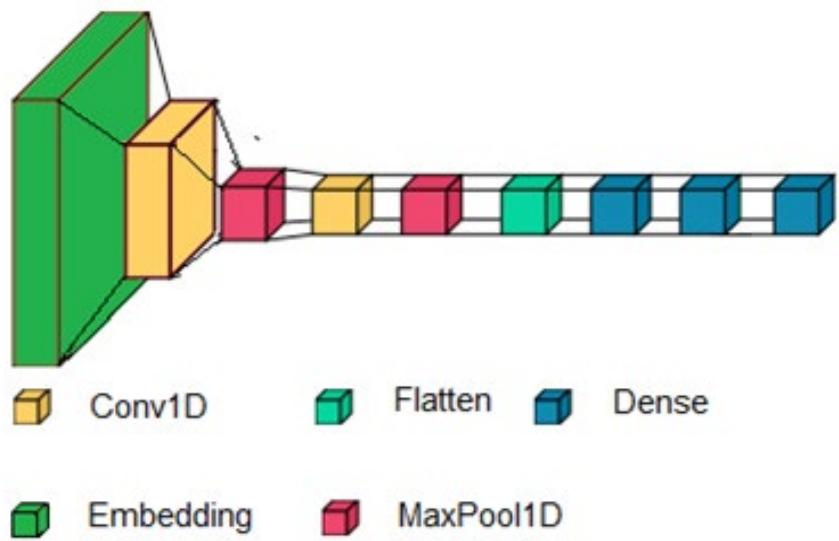

**Figure 3: Stage1 Model Architecture**

### Stage2 (Multiclass Classification Model)
The network consists of two 1-D convolution layers, two LSTM layers, and two fully connected layers. The input to the network is a preprocessed text, where each word embedded into 300-D vector, and the total number of words are truncated into 400 words. The number of layers was selected so as to maintain a high level of accuracy while still being fast enough for real-time purposes. In addition, it utilized max pooling and batch normalization more effectively in order to minimize overfitting (see Figure 4 Stage2 Model Architecture).

## Results & Discussion
### Stage1 Results

| Dataset | Model | Accuracy | | Hyper parameters | | | | |
|---|---|---|---|---|---|---|---|---|
| | | Train | Test | Optimizer | Batch | Epoch | Learning Rate | Activation function |
| MSR [16] | Convolution + LSTM | 96% | 95% | Adam | 64 | 15 | 0.001 | ReLU |
| Chrome & Debian [17] | Convolution + LSTM | 93% | 93% | Adam | 32 | 22 | 0.001 | ReLU |
| SySeVR [7] | Convolution + LSTM | 97% | 97% | Adam | 32 | 50 | 0.001 | ReLU |
| | Convolution | 74% | 74% | RMSprop | 128 | 10 | 0.005 | Tanh |
| | | 74% | 74% | Adam | 32 | 10 | 0.001 | Tanh |
| | | 74% | 74% | Adagrad | 16 | 10 | 0.0001 | ReLU |
| | | 99% | 98% | Adam | 64 | 10 | 0.005 | ReLU |
| | | 99% | 99% | Adam | 64 | 100 | 0.09 | ReLU |

**Table 4: Binary Classification Model Experiments.**

As shown in the table, the best accuracy that of Stage1 is 99% (see Figure 5 Stage1 Accuracy) on both training and testing sets using neural network architecture consisting of 2 convolution layers with 256 and 128 kernels for each, and the kernel size is 7, and ReLU was used as an activation function, followed by a fully connected neural network with 3 dense layers that contain 64,16 hidden neurons in each of them, the last layer contains 1 neuron where the activation function is Tanh to make the classification (see Stage1 (Binary classification model)). Since the output of this stage model is a value between 0 and 1 then Binary Cross-Entropy (BCE) was applied to all the experiments to compare each of the anticipated probabilities to the actual class output, which might either be 0 or 1. The following equation represents it mathematically [14]:

$$BCE(t,p) = -(t* \log(p) + (1-t)*\log(1-p))$$

### Stage2 Results

| Dataset | Model | Accuracy | | Hyper parameters | | | | |
|---|---|---|---|---|---|---|---|---|
| | | Train | Test | Optimizer | Batch | Epoch | Learning Rate | Activation function |
| MSR [16] | Convolution + LSTM | 74% | 48% | Adam | 32 | 100 | 0.001 | ReLU |
| SySeVR [7] | Convolution + LSTM | 93% | 93% | Adam | 32 | 22 | 0.001 | ReLU |
| | | 76% | 76% | Adagrad | 64 | 10 | 0.001 | Tanh |
| | | 87% | 86% | Adam | 128 | 20 | 0.001 | ReLU |
| | | 95% | 95% | RMSprop | 16 | 10 | 0.001 | ReLU |
| | | 97% | 96% | Adam | 32 | 50 | 0.0005 | ReLU |
| | | 98% | 98% | Adam | 32 | 50 | 0.001 | ReLU |
| | | 99% | 99% | Adam | 64 | 100 | 0.09 | ReLU |

**Table 5: Multiclass Classification Model Experiments.**

Stage2 achieves an accuracy of 98% (see Figure 6 Stage2 Accuracy) using a neural network consisting of 2 convolutions where the number of kernels is 64 and 128 with a kernel size of 3 with ReLU activation function for each, followed by 2 LSTMs of 100,10 cells with Tanh activation function, and 2 dense layers at the end of the neural network each have 100,50 neurons and the activation function is sigmoid to compute the probability of each class among the sample (see Stage2 (Multiclass classification model)). Since this stage performs a multiclass classification task then the cross-entropy loss function is used, which is a measure of the difference between two probability distributions for a given random variable or set of events. There are two types of cross entropy one for binary label and the other for the multiclass label. The following equation represents it mathematically [14]:

$$CCE(p,t) = -\sum_{c=1}^{C} t_{o,c} \log(P_{o,c})$$

| Paper | Dataset | Model | Accuracy |
|---|---|---|---|
| Deep Learning based Vulnerability Detection: Are We There Yet? [12] | ReVeal [8] | GGNN | 84% |
| Security Vulnerability Detection Using Deep Learning Natural Language Processing [19] | NVD/SARD [20] | BERT+BLSTM | 93% |
| SySeVR: A Framework for Using Deep Learning to Detect Software Vulnerabilities [10] | SySeVR [7] | BGRU | 96% |
| Automated Vulnerability Detection in Source Code Using Deep Representation Learning [13] | Draper VDISC [9] | CNN+RF | 91.6% |
| Combining Graph Neural Networks with Expert Knowledge for Smart Constract Vulnerablility Detection [11] | ESC and VSC [5] | GCE | 89% |
| Multi-context Attention Fusion Neural Network for Software Vulnerability Identification [21] | SARD [22] | Attention Fusion Model | 99% |
| Our Proposed Solution | SySeVR [7] | Convolutional | 99.2% |
|  |  | Convolutional + LSTM | 98.7% |

**Table 6:** Comparison with Previous Works.

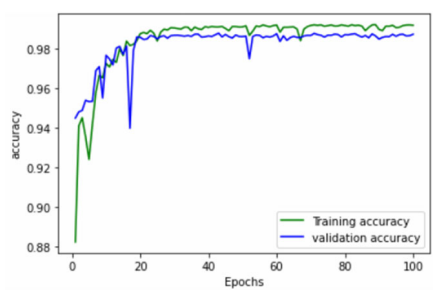

**Figure 5:** Stage1 Accuracy

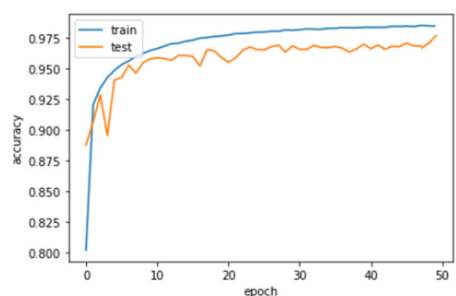

**Figure 6:** Stage2 Accuracy

## Conclusion & Future Work

New vulnerabilities appear every day and cause attacks. A two-stage approach for detecting vulnerabilities in C/C++ source codes was proposed in this research, each stage is a deep learning model. The first stage looks for a vulnerability in the source code, and the second stage categorizes it. Results show that two-stages solution outperforms single-stage solutions. Also, embedding codes outperform extracting features from them. We aim to improve our solution by modeling the source code as a graph and feeding it into a graph neural network to detect vulnerabilities, as well as detecting multiple vulnerabilities, detecting the position of the vulnerability, and finally using the sequence-to-sequence language model to fix the vulnerable code to make it non-vulnerable [15-22].

## Declarations
### Authors' Contributions
MAH took on the main role so he performed the literature review, conducted the experiments, and wrote the manuscript.

MH performed the literature review and conducted the experiments.

KAJ took on a supervisory role and contributed to the conception and analysis of the work. All authors read and approved the final manuscript.

All authors read and approved the final manuscript.

### Funding
The authors declare that they have no funding.

### Availability of Data and Materials
The dataset used in this work is available online, you can find its link in references.

### Ethics Approval and Consent to Participate
The authors Ethics approval and consent to participate.

### Consent for Publication
The authors consent for publication.

### Competing Interests
The authors declare that they have no competing interests.

## References
1. Black, P. E. (2018). A software assurance reference dataset: Thousands of programs with known bugs. Journal of research of the National Institute of Standards and Technology, 123, 1.
2. Li, H., & Zhang, X. (2011). A Building Method of Virtual Knowledge Base Based on Ontology Mapping. In

Knowledge Engineering and Management: Proceedings of the Sixth International Conference on Intelligent Systems and Knowledge Engineering, Shanghai, China, Dec 2011 (ISKE2011) (pp. 597-602). Springer Berlin Heidelberg.
3. Bhandari, G., Naseer, A., & Moonen, L. (2021, August). CVEfixes: automated collection of vulnerabilities and their fixes from open-source software. In Proceedings of the 17th International Conference on Predictive Models and Data Analytics in Software Engineering (pp. 30-39).
4. Karampatsis and C. Sutton, "zenodo," 7 2 2020. [Accessed 20 7 2020].
5. Messi-Q, "Github," 16 4 2020. [Accessed 20 7 2022].
6. vntchain, "Github," 22 6 2020. [Accessed 20 7 2022].
7. SySeVR, "GitHub," 18 7 2018. [Accessed 20 7 2022].
8. Liu, Z., Qian, P., Wang, X., Zhuang, Y., Qiu, L., & Wang, X. (2021). Combining graph neural networks with expert knowledge for smart contract vulnerability detection. IEEE Transactions on Knowledge and Data Engineering.
9. Chakraborty, S., Krishna, R., Ding, Y., & Ray, B. (2021). Deep learning based vulnerability detection: Are we there yet. IEEE Transactions on Software Engineering.
10. Russell, R., Kim, L., Hamilton, L., Lazovich, T., Harer, J., Ozdemir, O., ... & McConley, M. (2018, December). Automated vulnerability detection in source code using deep representation learning. In 2018 17th IEEE international conference on machine learning and applications (ICMLA) (pp. 757-762). IEEE.
11. Ma, Y., & He, H. (Eds.). (2013). Imbalanced learning: foundations, algorithms, and applications.
12. "CWE List," cybersecurity-help, [Accessed 20 7 2022].
13. Fan, J., Li, Y., Wang, S., & Nguyen, T. N. (2020, June). AC/C++ code vulnerability dataset with code changes and CVE summaries. In Proceedings of the 17th International Conference on Mining Software Repositories (pp. 508-512).
14. "Loss Functions," ML Glossary, [Accessed 20 7 2022].
15. S. Chakraborty, "Google Drive," 2 5 2020.
16. Ziems, N., & Wu, S. (2021, May). Security vulnerability detection using deep learning natural language processing. In IEEE INFOCOM 2021-IEEE Conference on Computer Communications Workshops (INFOCOM WKSHPS) (pp. 1-6). IEEE.
17. [Online]. Available: NATIONAL VULNERABILITY DATABASE.
18. Tanwar, A., Manikandan, H., Sundaresan, K., Ganesan, P., Chandrasekaran, S. K., & Ravi, S. (2021). Multi-context Attention Fusion Neural Network for Software Vulnerability Identification. arXiv preprint arXiv:2104.09225.
19. "NIST SOFTWARE ASSURANCE REFERENCE DATASET," National Institute of Standards and Technology, [Accessed 20 7 2022].
20. VulDetProject, "Github," 31 5 2020. [Accessed 20 7 2022].
21. L. Kim and R. Russell, "OSF HOME," 20 11 2018. [Accessed 20 7 2022].
22. Li, Z., Zou, D., Xu, S., Jin, H., Zhu, Y., & Chen, Z. (2021). Sysevr: A framework for using deep learning to detect software vulnerabilities. IEEE Transactions on Dependable and Secure Computing, 19(4), 2244-2258.